# Third Harmonic Susceptibility and the Irreversibility Line of $Fe/MgB_2$ Tapes

C. Senatore [1,2], N. Clayton [2], P. Lezza [2], M. Polichetti [1], S. Pace [1], R. Flükiger [2]

[1]Dipartimento di Fisica "E.R.Caianiello" and INFM, Università degli Studi di Salerno, Baronissi, Italy
[2]Département de Physique de la Matière Condensée, Université de Genève, Genève, Switzerland

## Abstract

We report a study of the pinning properties of $Fe/MgB_2$ tapes performed by means of ac magnetic susceptibility measurements. In particular, the third harmonic response has been measured as a function of the dc magnetic field up to 9 T, at different temperatures, for various frequencies and amplitudes of the ac field. The irreversibility line has been determined using the third harmonic onset criterion for samples made with different ball-milled $MgB_2$ powders. The study of the harmonic response allows us to probe the nature of the vortex pinning and to analyse its influence on the enhancement of the irreversibility field in strongly ball-milled samples. We have also investigated the dynamical regimes governing the vortex motion in the tapes by comparing the experimental curves with numerical simulations of the non-linear diffusion equation for the magnetic field.

*Contacting author:* Carmine Senatore (e-mail: Carmine.Senatore@Physics.Unisa.it)

## 1. Introduction

Since the discovery of superconductivity in $MgB_2$ below a critical temperature $T_c$ of 39 K [1], significant progress has been made in order to understand and develop this material to exploit its high intrinsic performance for power applications, such as magnets and transformers. It is believed that the supercurrent density in $MgB_2$ is controlled predominantly by flux pinning rather than by grain boundary connectivity because of the large coherence length in this material ($x_0$ = 50 Å) [2]. However, the lack of natural defects is responsible for the rapid decline of the critical current density with increasing magnetic field strength. In view of the applications mentioned above, soon after the discovery of $MgB_2$, many research groups tried to demonstrate the feasibility of fabrication of wires and tapes and several processing methods have been developed in order to obtain high critical current densities [3]. The most studied approach has been the power-in-tube (PIT) process, where powder is packed into a metal tube and drawn into a wire. The wire can subsequently be rolled to form a tape. Indeed, very high critical current densities ($J_c$ » $10^5$÷$10^6$ A/cm$^2$) have been reported at 5 K [4-6], but these high values rapidly decrease with increasing temperature and magnetic field. To improve the performance of $MgB_2$ wires, it is necessary to find a processing method that can introduce more pinning centres and also overcome the poor connectivity between the powder grains, which is mainly caused by porosity. In order to do this, it is essential to understand the flux pinning mechanism and the irreversibility properties of $MgB_2$.

The third harmonic response of the ac magnetic susceptibility is a highly sensitive tool to investigate the flux dynamics in relation to the electrical transport properties of the sample [7,8]. In the Bean critical state model the harmonic generation is attributed to the hysteretic relationship between the magnetization and the external field due to flux pinning [9]. However, in order to account for the frequency dependence of the ac magnetic response, the simultaneous presence of hysteretic and dynamic losses has to be included in the model. In fact, when the vortex motion is governed by thermally activated flux creep, the *I*-*V* curve of a type-II superconductor exhibits non-linear behaviour, which is a direct consequence of the current dependence of the pinning energy *U*, and this leads to the generation of higher harmonics in the ac susceptibility [8].

In this paper we investigate the effect of ball-milling the $MgB_2$ powders, and the consequent reduction of the particle size, on the pinning properties of monofilamentary Fe/$MgB_2$ tapes by measuring the third harmonics ($c_3 = c'_3 + ic''_3$) of the ac magnetic susceptibility as a function of the dc magnetic field, for various frequencies and amplitude of the ac field. We have determined the irreversibility line by using the third harmonic onset criterion. In particular, we have studied the nature of the enhancement of the irreversibility field, due to the ball-milling process, by performing numerical simulations of the non-linear magnetic diffusion equation.

The paper is organized as follows. The sample preparation and the experimental procedure are described in Sec.2. In Sec.3 we report the numerical method applied to solve the magnetic diffusion equation. We present in Sec.4 the $c_3(B)$ measurements and the irreversibility line for Fe/$MgB_2$ tapes prepared with different ball-milling times. Finally, Sec.5 reports the conclusions of this work.

## 2. Experimental details

The Fe/$MgB_2$ tapes studied in this paper have been fabricated by the PIT technique. $MgB_2$ powder has been introduced into Fe tubes inside a glove box under an Ar atmosphere, which have then been swaged, drawn and rolled. The Fe tubes used in this experiment had an outer diameter of 8 mm and an inner diameter of 5 mm. Both ends of the tube were sealed with lead pieces. After swaging to 3.85 mm and drawing to 2 mm, the wire is deformed by flat rolling to obtain a $3.9\times0.38$ mm$^2$ tape. A 920°C final heat treatment is performed for 0.5 h in a pure Ar atmosphere.

In this work we report measurements performed on three different tapes in order to study the effect of the initial powder size on the pinning properties. One tape has been prepared directly from the as-purchased powder, i.e. standard commercial $MgB_2$ powder from Alpha-Aesar with a purity of 98%. Two other tapes have been prepared by ball-milling the starting powder for 3h and 100h under Ar atmosphere. Granulometry measurements show that the as-purchased powder contains a large number of agglomerated grains with a wide size distribution centred at around 60 µm, and a maximum diameter of about 120 µm. After the ball-milling process the maximum diameter of the grains is around 20 µm, while in the 3h /100h ball-milled powder 35% / 50% of the grains are of sub-micron size.

The ac susceptibility measurements reported in this work have been performed on the $MgB_2$ cores of three tapes using an home made susceptometer. The magnetic field dependence of both the first and the third harmonics ($c_1(B)$ and $c_3(B)$) has been measured by sweeping the dc field from 0 to 9 T, at different temperatures (15, 20, 25, 30, 35 K). Measurements have been performed using different ac field amplitudes (1, 2, 4, 8, 16 G) in the frequency range 1007 ÷ 5007 Hz. Both the ac and dc fields are applied parallel to the longest side of the sample.

Because of the sensitivity of the third harmonic response to the phase setting of the measurements, we have set the phase in a very accurate way. We have measured the wide-band susceptibility [10] as a function of the lock-in phase in the range [-180°, 180°] for each measured frequency at $B_{ac}$ = 1 G and $T$ = 4.2 K, both with and without the sample. We then calculate the difference between the curves acquired with and without the sample in order to remove all the contributions due to spurious signals. The amplitude susceptibility $c_a$ is proportional to the sample magnetization when the external field reaches its maximum value, whereas the remanent susceptibility $c_r$ is proportional to the magnetization remaining in the sample at the zero instantaneous value of the ac field. When the sample is in the Meissner state, $c_r$ is zero and $c_a$ reaches its maximum negative value. This allows us to determine the phase to an accuracy better than 0.1 degree.

## 3. Numerical method

In order to relate the harmonic susceptibilities to the vortex pinning and dynamics, we have numerically solved the non-linear diffusion equation which governs the magnetic flux penetration in a type-II superconductor. In the case of an infinite slab of thickness $d$ in a parallel field geometry this equation, derived by the Maxwell equations and Ohm's law, can be written as:

$$\frac{\partial B}{\partial t} = \frac{\partial}{\partial x}\left[\frac{r(B,J,T)}{m}\frac{\partial B}{\partial x}\right] \tag{1}$$

with the boundary conditions $B(\pm d/2,t) = B_{ext}(t) = B_{dc} + B_{ac}\, sin(2p\ f\ t)$. The diffusion coefficient $r\ (B,J,T)$ is the residual resistivity due to thermally activated flux motion. In fact this leads to an electric field, whose expression is:

$$E = r_n \frac{B}{B_{c2}(T)} J_c(B,T) exp\left[-\frac{U_P(B,T,J)}{k_B T}\right]. \tag{2}$$

We assume that the temperature, magnetic field and current density dependence in the effective energy barrier $U_P(B, T, J)$ may be separated, using for the current dependence the Anderson-Kim expression:

$$U_P(B,J,T) = U_0(B,T)\left[1 - \frac{J}{J_c(B,T)}\right]. \tag{3}$$

This expression has been derived by assuming that flux creep occurs by bundles of flux lines jumping between adjacent pinning points [11].

Equation (1) has been numerically solved by means of Fortran NAG [12] routines. In order to obtain the harmonic susceptibilities $c_n = c'_n + ic''_n$, we have first calculated the magnetization loop for the applied time-dependent field:

$$m_0 M(t) = \frac{1}{d}\int_0^d B(x)\,dx - \left[B_{dc} + B_{ac}\, sin(2p\ f\ t)\right]\ , \tag{4}$$

and then its Fourier transforms:

$$c'_n = \frac{1}{pB_{ac}}\int_0^{2p} m_0 M(t)\, sin(nw\,t)\, d(w\,t) \tag{5a}$$

$$c''_n = \frac{1}{pB_{ac}}\int_0^{2p} m_0 M(t)\, cos(nw\,t)\, d(w\,t) \tag{5b}$$

To account for the magnetic field and temperature dependence of the ac susceptibilities we have to specify the magnetic field and temperature dependence of the critical current density $J_c(B,T)$ and the thermal activation energy $U_0(B,T)$. To this end, we chose the following forms [13,14]:

$$J_c(B,T) = J_c(B=0,T=0)\frac{B_1}{B_1+B}\frac{\left(1-t^2\right)^{\frac{5}{2}}}{\left(1+t^2\right)^{\frac{1}{2}}} \tag{6a}$$

$$U_0(B,T) = U_0(B=0,T=0)\frac{B_2}{B_2+B}\left(1-t^4\right), \tag{6b}$$

where $t$ represents the reduced temperature $T/T_c$ and the values of the parameters in the magnetic field dependence are $B_1 = 2.5\ 10^{-2}$ T and $B_2 = 5\ 10^{-2}$T. The upper critical field is [15]:

$$B_{c2}(T) = B_{c2}(T=0)\ \frac{1-t^2}{1+t^2}, \tag{7}$$

with $B_{c2}(T=0) = 20$ T .

The numerical calculations of the third harmonic response reported in this paper have been performed as a function of the ratio $U_0\,(B=0,T=0)/\,k_BT_c$ for fixed $J_c\,(B=0,T=0)\times d$, simulating an enhancement of the pinning and as a function of the value of $J_c\,(B=0,T=0)\times d$ for fixed $U_0\,(B=0,T=0)/\,k_BT_c$, simulating an improvement of the grain connectivity. This has been done in order to investigate the modification induced in the vortex dynamics by means of the ball-milling process, as will be shown in the next section.

## 4. Experimental results and discussion

In Fig.1 we report the magnetic field dependence of the real and imaginary components of the first harmonic $c_1$ (Fig.1a) and the third harmonic $c_3$ (Fig.1b) measured at $T$ =20 K on the 3 h ball-milled tape with $f$ =1607 Hz and $B_{ac}$ =16 G. According to the critical state prediction, reported in the inset, the real part of the first harmonic $c'_1$ shows the superconducting transition whereas the imaginary part exhibits a peak that corresponds to the losses of the system. We can also observe the typical structures predicted by the Bean model, i.e. a peak that rises when the relation $B_{ac}=J_c(B_{dc})\times d$ holds and goes monotonously to zero in $c'_3(B)$ and a peak followed by a dip in $c''_3(B)$. We note that the height of the $c_3(B)$ peaks is higher in the experimental curves than in the curves calculated in the framework of the critical state. Moreover, in the Bean model the peak in the imaginary part $c''_3(B)$ is predicted to be higher than the one in the real part $c'_3(B)$ (inset Fig.1b), whereas in the experiment the opposite is observed. This behaviour can be ascribed to flux dynamics. In particular, since the flux flow regime becomes linear when $B_{dc} >> B_{ac}$ , the only dynamic regime contributing to the signal of higher harmonics is flux creep. Similar features are encountered in the measurements performed on the other tapes.

Fig.2 shows the magnetic field dependence of the third harmonic modulus $|c_3|$ measured at different temperatures ($T$ =15, 20, 25, 30 K) on the non-ball-milled tape, with $f$ =1607 Hz and $B_{ac}$ =4 G. The curves exhibit a pronounced granular behaviour, due to the porosity of the sample. Because of the poor connectivity between the grains we can observe in the $|c_3|(B)$ curves two well separated peaks. The low field peak is determined by intergranular currents, likely due to weak Josephson coupling. At higher fields, the intergranular currents become negligible and the behaviour of the sample is governed by the decoupled grains: the resulting high field peak corresponds to the currents flowing inside the grains. In Fig.3 the magnetization loop measured on the same sample at 5 K for a sweep rate of 2 T/min. is reported; the measurement has been performed using a Vibrating Sample Magnetometer (VSM). The loop exhibits the typical shape due to flux pinning on a surface barrier and this further confirms the granular behaviour of the sample. In fact, because of the low electrical field induced in the sample in a VSM measurement (typically $(dB/dt)_{VSM}/(dB/dt)_c$ » $10^{-1}$÷$10^{-2}$), the grains are completely decoupled. Moreover, the grain surfaces acts as surface barriers and the lack of strong bulk pinning determines the Z-shape of the hysteresis loop.

Fig.4 shows the irreversibility line of the tapes, whose powders have had different ball-milling times, determined by using the third harmonic onset criterion. In the presence of an applied magnetic field $B_{dc} >> B_{ac}$ the onset of the third harmonic is the result of a dynamic crossover from a regime in which the system's response is dominated by flux flow, and characterized by the absence of harmonic signal to one that is dominated by pinning. Therefore, the onset field corresponds to the irreversibility field, as defined in the flux creep model [16].

The irreversibility line of the non-ball-milled tape in Fig.4 reflects the intragrain peak of the $c_3$ $(B)$ curves. We notice that, as a consequence of the ball-milling process, the irreversibility field is strongly enhanced. In particular, the value $B_{irr}$ =9 T is reached at $T$ =15 K for the non-ball-milled tape, at $T$ =16 K for the 3h ball-milled tape, at $T$ =18 K for the 100h ball-milled tape.

The ball-milling process reduces the size of the powder grains and this leads to an improvement of the connectivity. The transport properties of the Fe/$MgB_2$ tapes are determined by both the strength of the pinning centres and the connectivity of the grains. In order to study the mechanism responsible for the enhancement of the irreversibility field in the ball-milled samples we performed numerical simulations changing, separately, the $J_c\,(B=0,T=0)\times d$ value and the $U_0\,(B=0,T=0)/\,k_BT_c$ ratio. The $|c_3|(B)$ curves calculated for $U_0\,(B=0,T=0)/\,k_BT_c$ =750, 1125, 1500 at fixed $J_c\,(B=0,T=0)\times d$ =5 $10^5$ A/m are plotted in Fig.5. The onset of $|c_3|$ and, thus, $B_{irr}$ increases with $U_0\,(B=0,T=0)/\,k_BT_c$, i.e. with the strength of the pinning. Moreover, the increase of $U_0\,(B=0,T=0)/\,k_BT_c$ determines the broadening of the peak and the increase of its height. On the other hand, when $J_c\,(B=0,T=0)\times d$ is increased at fixed $U_0\,(B=0,T=0)/\,k_BT_c$, simulating an increase of the current density due to the improvement of the connectivity and but not of pinning, the onset of $|c_3|$ is unaffected while the peak becomes sharper, and its height is reduced (Fig.6).

In Fig.7 we report the $|c_3|(B)$ curves measured at $T$ =20 K with $B_{ac}$ =8 G and $f$ =1607 Hz for the three different tapes. By comparing the experimental curves with the simulated ones we can associate the shift of the third harmonic onset to higher fields and, thus, the increase of $B_{irr}$ to the enhancement of the pinning. Since the upper critical field $B_{c2}$ and the critical temperature $T_c$ have been found in the literature to be unaffected by the ball-milling process [17,18], this enhanced pinning is likely to be confined to the grain surface [3]. Nevertheless the disappearance of the intragrain peak in the 3h and 100h ball-milled curves can be ascribed to an improvement of the grain connectivity due to the ball-milling process.

In Fig.8 we report the $|c_3|(B)$ curves measured at fixed temperature and ac field ($T$ =20 K, $B_{ac}$ =4 G) for two frequencies ($f$ = 1007 and 5007 Hz). The frequency behaviour of the third harmonic response is related to the current density dependence of the pinning energy $U_P(B, T, J)$ and, thus, to the pinning mechanisms [19,20]. Measurements have been performed on four different samples, namely the commercial powders (Fig.8a), the non-ball-milled tape (Fig.8b), the 3h ball-milled tape (Fig.8c), the 100h ball-milled tape (Fig.8d). The interesting feature is the increase of the $|c_3|(B)$ peak amplitude with the frequency, common to all the samples. This means that the ball-milling of the initial powder, the deformations during the fabrication of the tape, the annealing process and the consequent recrystallization do not change the pinning mechanism present in the powders, but lead to the an increase of the strength of the pinning centres. Moreover, we reproduce the frequency behaviour of the $|c_3|(B)$ curves in the numerical simulations (Fig.9) by using the Anderson-Kim pinning model, corresponding, as shown in Sec.3, to the dependence $U(J)\propto 1-(J/J_c)$. The parameters used for the simulated $|c_3|(B)$ curves in Fig.9 are $U_0\,(B=0,T=0)/\,k_BT_c=1125$, $J_c\,(B=0,T=0)\times d=5\times10^5$ A/m, $T/T_c=0.5$, $B_{ac}=4$ G, $f$ = 1007 and 5007 Hz. The Anderson-Kim model implies the presence of extended pinning centres, i.e. whose dimensions are large compared with the inter-flux-line spacing ($=1.07(f_0/B)^{1/2}$), whereas collective pinning arises from point-size disorder randomly distributed in the superconductor [21]. The collective pinning mechanism ($U(J)\propto(J_c/J)^m$), that has been found to correctly describe the frequency behaviour of the third harmonic response for $MgB_2$ bulk samples [8,19], has to be excluded in this case since it leads to the decrease of the $|c_3|$ peak amplitude with the frequency, as reported in Ref.[19].

## 5. Conclusions

In this paper, the third harmonic susceptibilities as a function of the dc magnetic field have been measured on Fe/$MgB_2$ tapes made with different ball-milled powders. We have determined the irreversibility line by using the third harmonic onset criterion. In order to probe the nature of the enhancement of the irreversibility in the strongly ball-milled samples we have performed numerical simulations of the non-linear magnetic diffusion equation. The ball-milling process produces both the improvement of the connectivity of the grains and an enhancement of the pinning. The latter effect leads to the increase of the irreversibility field $B_{irr}$. Moreover, the frequency behaviour of the $|c_3|(B)$ curves suggests that the vortex dynamics in the measured samples can be described by the Anderson-Kim model, excluding the presence of the collective pinning mechanism. This demonstrates the extended size of the pinning centres.

**Aknowledgements**

We express our thanks to Dr. Bernd Seeber for useful suggestions. Thanks are also due to Mr. Robert Modoux and Mr. Alexandre Ferreira for their technical support.

## Figure captions

**Figure 1**: Magnetic field dependence of $c'_1$ and $c''_1$ in (a), $c'_3$ and $c''_3$ in (b) measured at $T$ =20 K, $f$ =1607 Hz and $B_{ac}$= 16 G on the 3h ball-milled sample. Insets: For comparison, $c'_1$ and $c''_1$ (a), $c'_3$ and $c''_3$ (b) as a function of the magnetic field, calculated in the critical state model.

**Figure 2**: Magnetic field dependence of $|c_3|$ measured at $f$ =1607 Hz, $B_{ac}$= 4 G and $T$ =15, 20, 25, 30 K on the non-ball-milled tape; the curves exhibit two well-separated peaks due to the granularity of the sample.

**Figure 3**: Magnetization loop for non-ball-milled tape measured at $T$ =5 K with a sweep rate of 2 T/min.; the loop shows the typical shape of flux pinning on a surface barrier.

**Figure 4**: Irreversibility line for the non-ball-milled tape (solid square), the 3h ball-milled tape (open circle), the 100h ball-milled tape (solid triangle); the curves have been determined by using the third harmonic onset criterion.

**Figure 5**: Magnetic field dependence of $|c_3|$ calculated for $T/T_c$= 0.5, $f$ =1607 Hz and $B_{ac}$ =8 G. The parameters used for the simulations are $J_c\,(B=0,T=0)\times\, d$ $=5\times10^5$ A/m and $U_0\,(B=0,T=0)/\,k_BT_c$ =750 (solid square), 1125 (open circle), 1500 (solid triangle); the onset of $|c_3|$ and, thus, $B_{irr}$ increases with $U_0\,(B=0,T=0)/\,k_BT_c$.

**Figure 6**: Magnetic field dependence of $|c_3|$ calculated for $T/T_c = 0.5$, $f$ =1607 Hz and $B_{ac}$ =8 G. The parameters used for the simulations are $U_0$ $(B=0,T=0)/\ k_BT_c$ =750 and $J_c$ $(B=0,T=0)\times d$ =2.5×10$^5$ A/m (solid square), 5×10$^5$ A/m (open circle), 7.5×10$^5$ A/m (solid triangle); the onset of $|c_3|$ does not change on increasing $J_c\,(B=0,T=0)\times d$.

**Figure 7**: Magnetic field dependence of $|c_3|$ measured at $T$ =20 K, $f$ =1607 Hz and $B_{ac}$ =8 G for the non-ball-milled tape (solid square), the 3h ball-milled tape (open circle), the 100h ball-milled tape (solid triangle); the shift of the onset to higher fields has to be ascribed to the enhancement of the pinning.

**Figure 8**: $|c_3|(B)$ curves measured at $T$ =20 K and $B_{ac}$ =4 G for two frequencies ($f$ = 1007 and 5007 Hz). Measurements have been performed on the commercial powders (a), the non-ball-milled tape (b), the 3h ball-milled tape (c), the 100h ball-milled tape (d); the height of the modulus peak increases as the frequency increases for all the samples.

**Figure 9**: Magnetic field dependence of $|c_3|$ calculated in the framework of the Anderson-Kim model for $T/T_c = 0.5$, $B_{ac}$ =4 G, $f$ =1007 Hz (solid square) and 5007 Hz (open square); the experimental behaviour in Fig.8 is well reproduced. The parameters used for the simulations are $U_0\,(B=0,T=0)/\ k_BT_c$ =750 and $J_c\,(B=0,T=0)\times d$ =5×10$^5$ A/m.

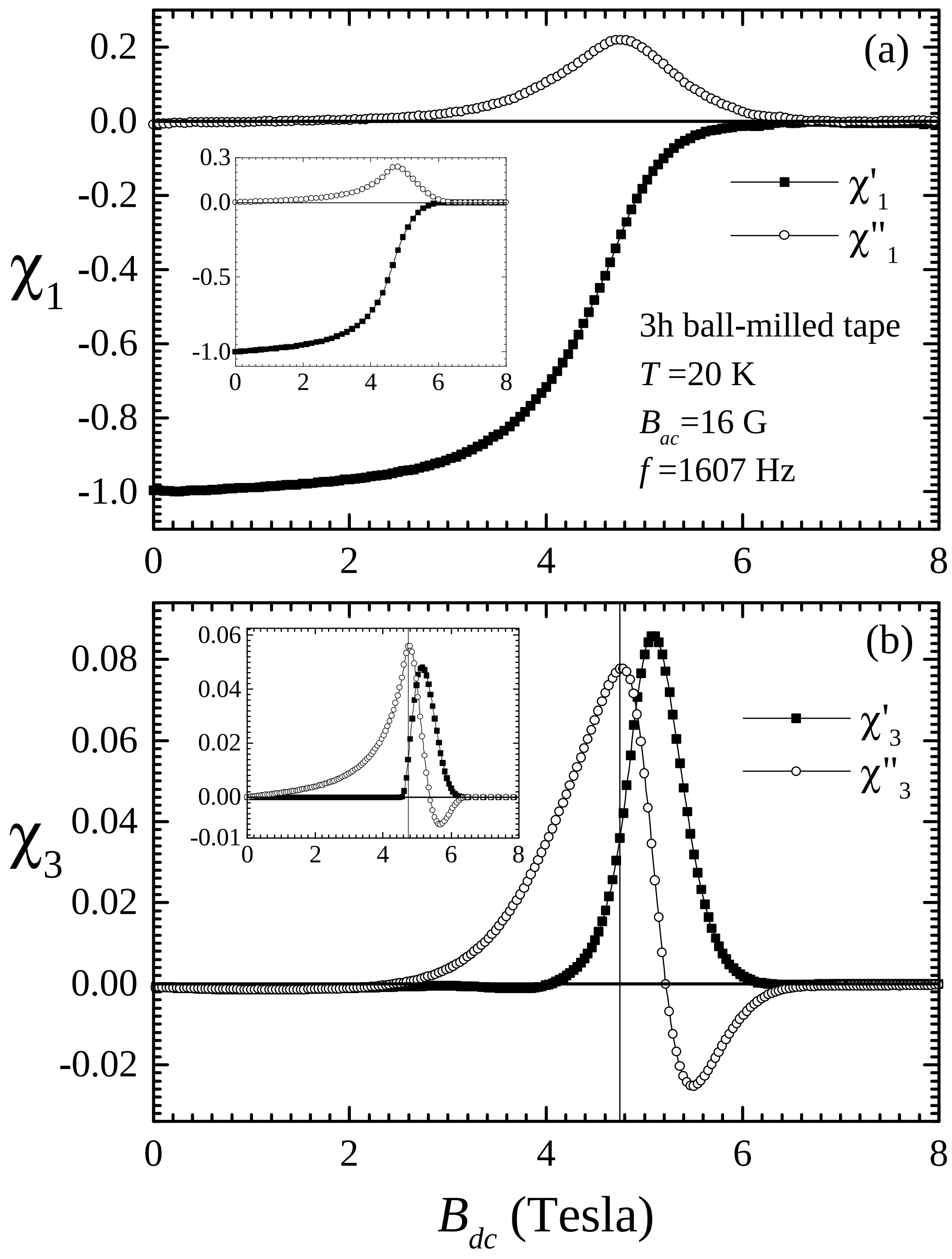

Figure 1

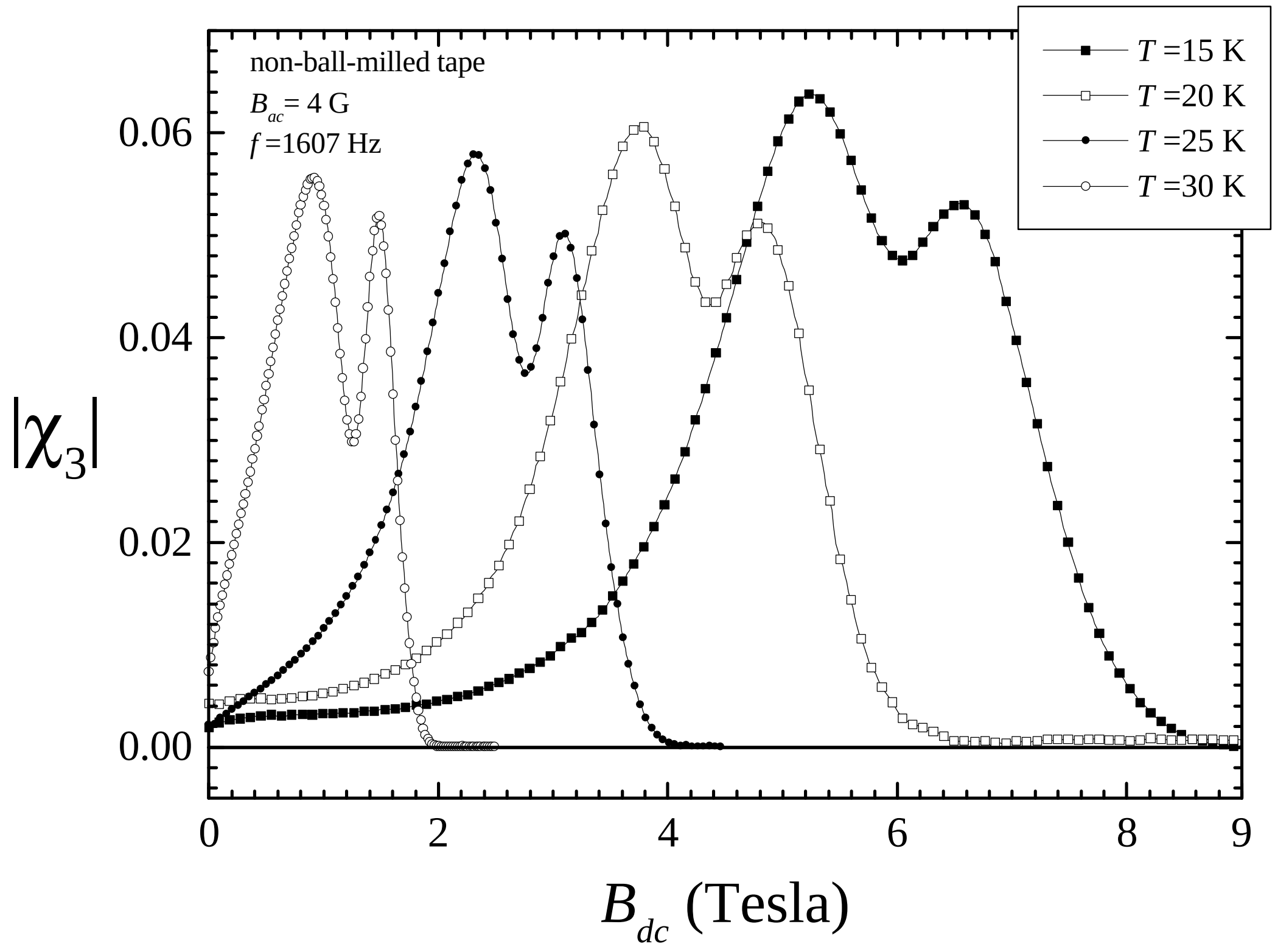

Figure 2

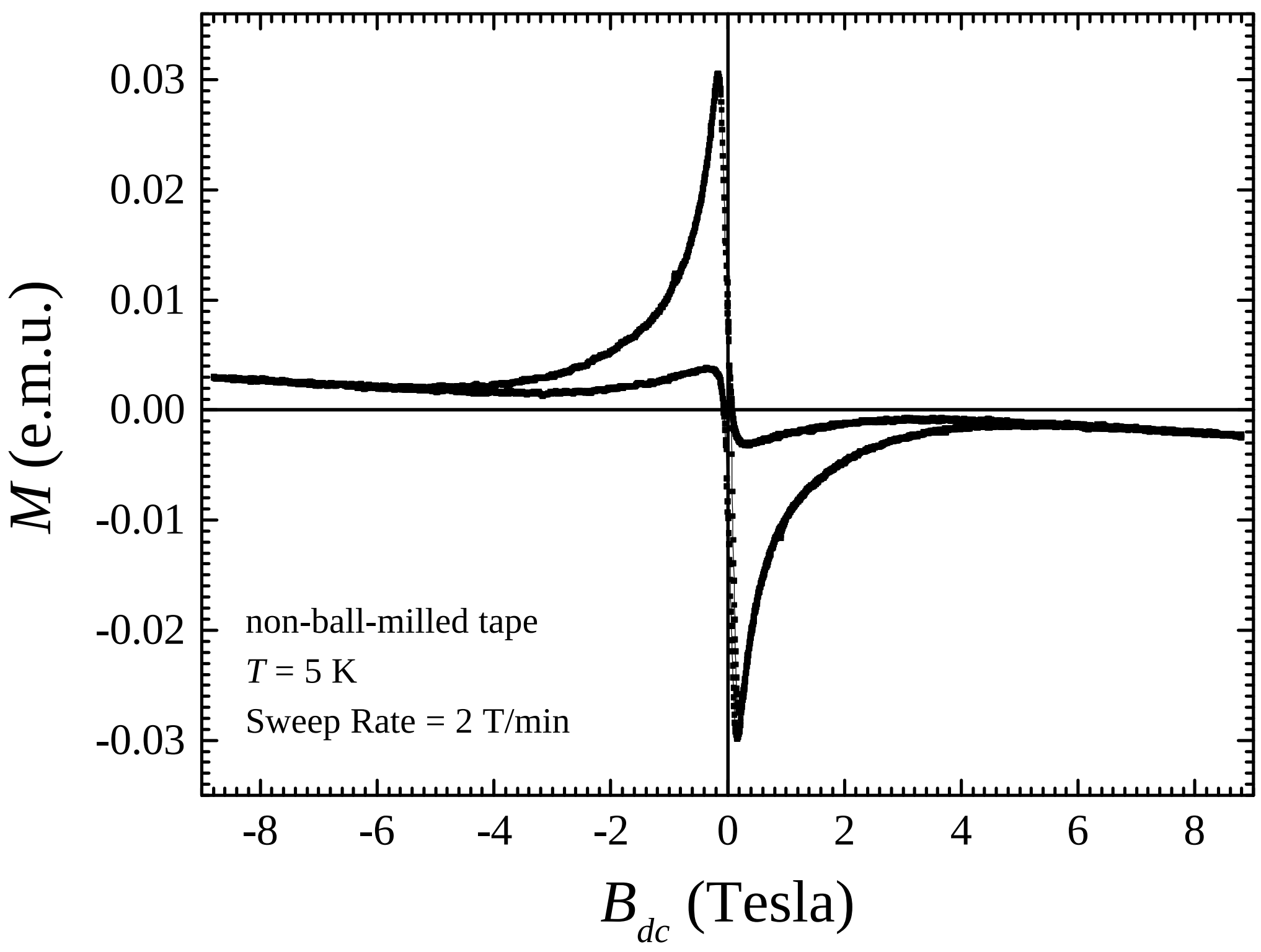

Figure 3

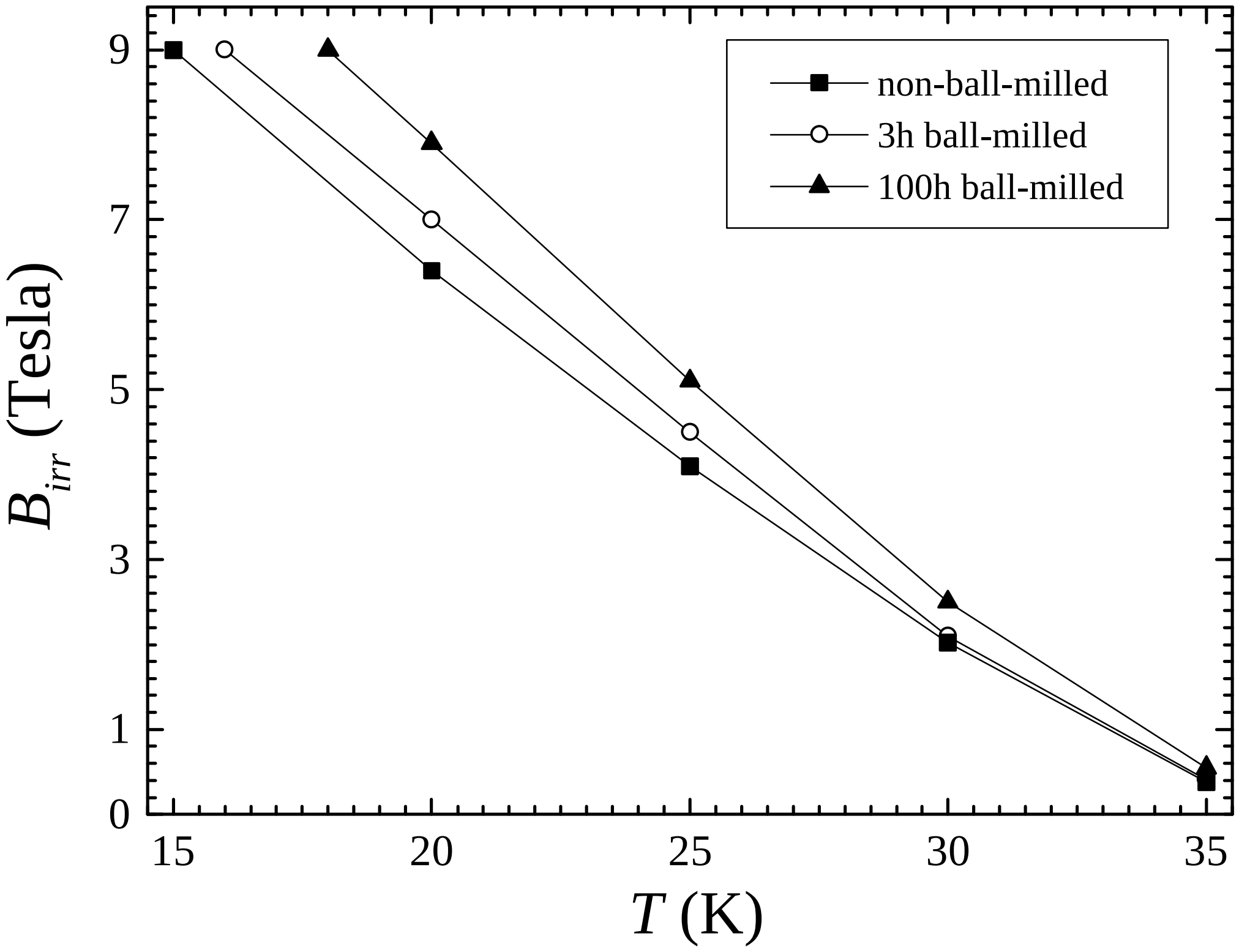

Figure 4

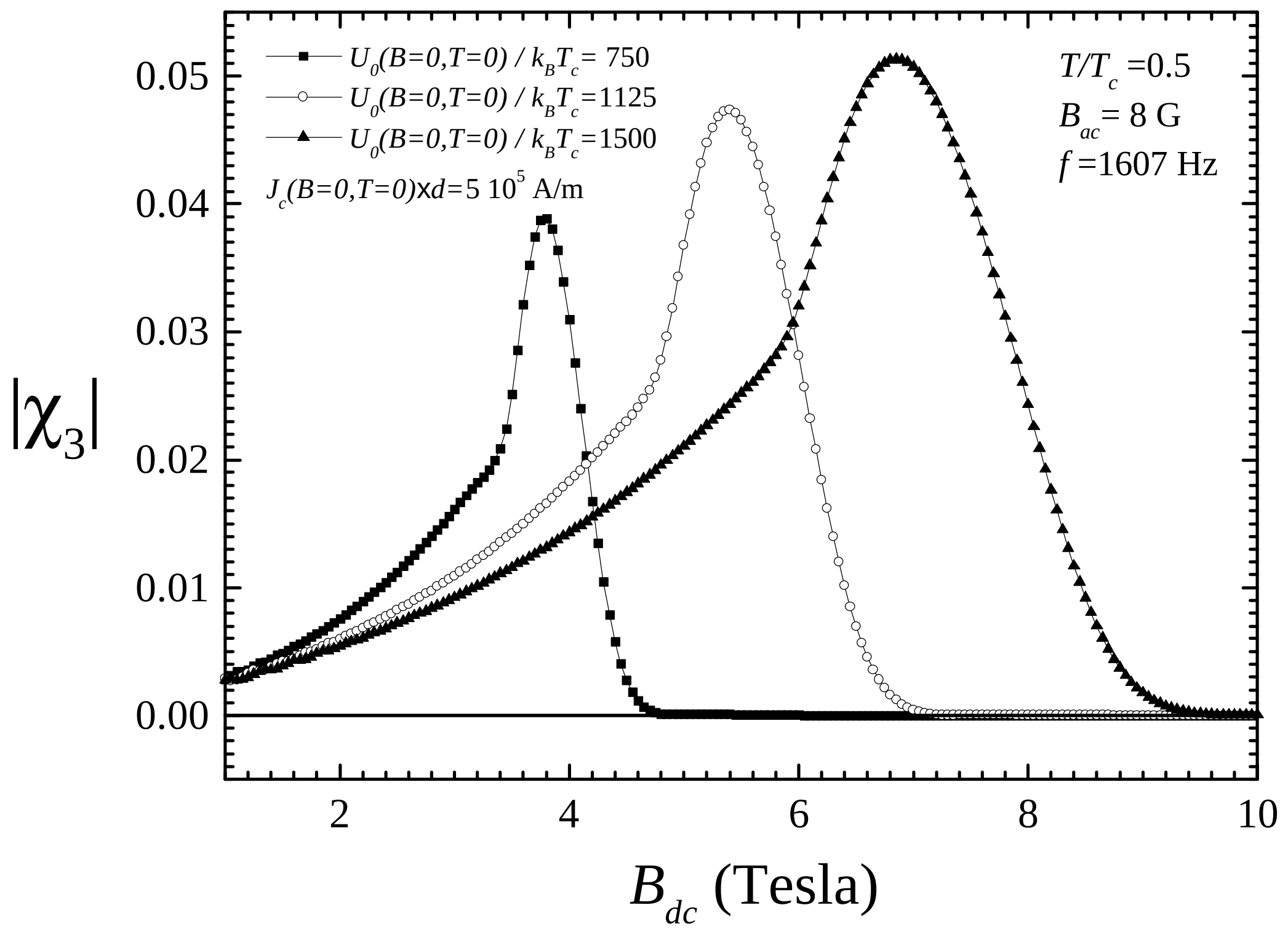

Figure 5

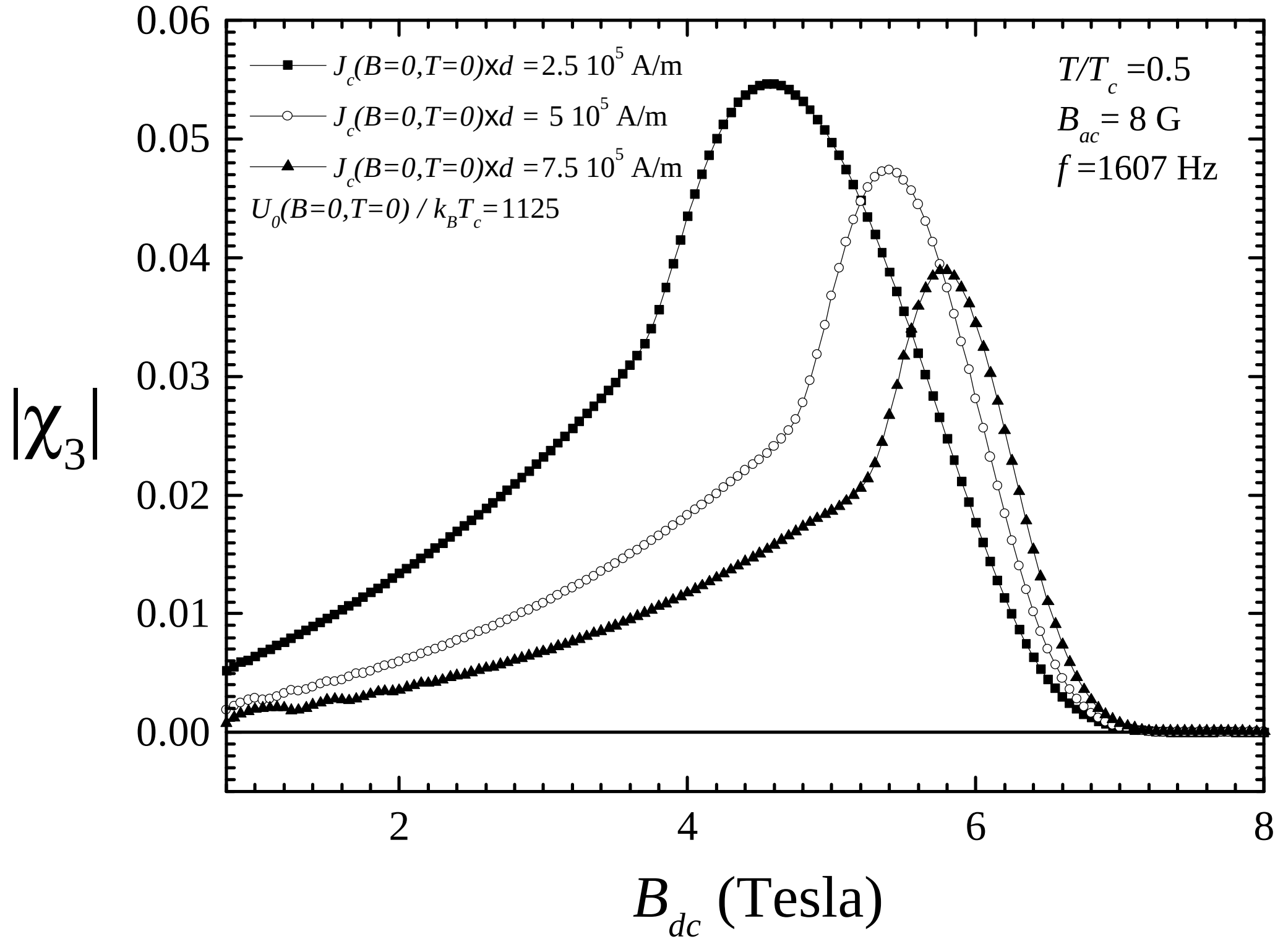

Figure 6

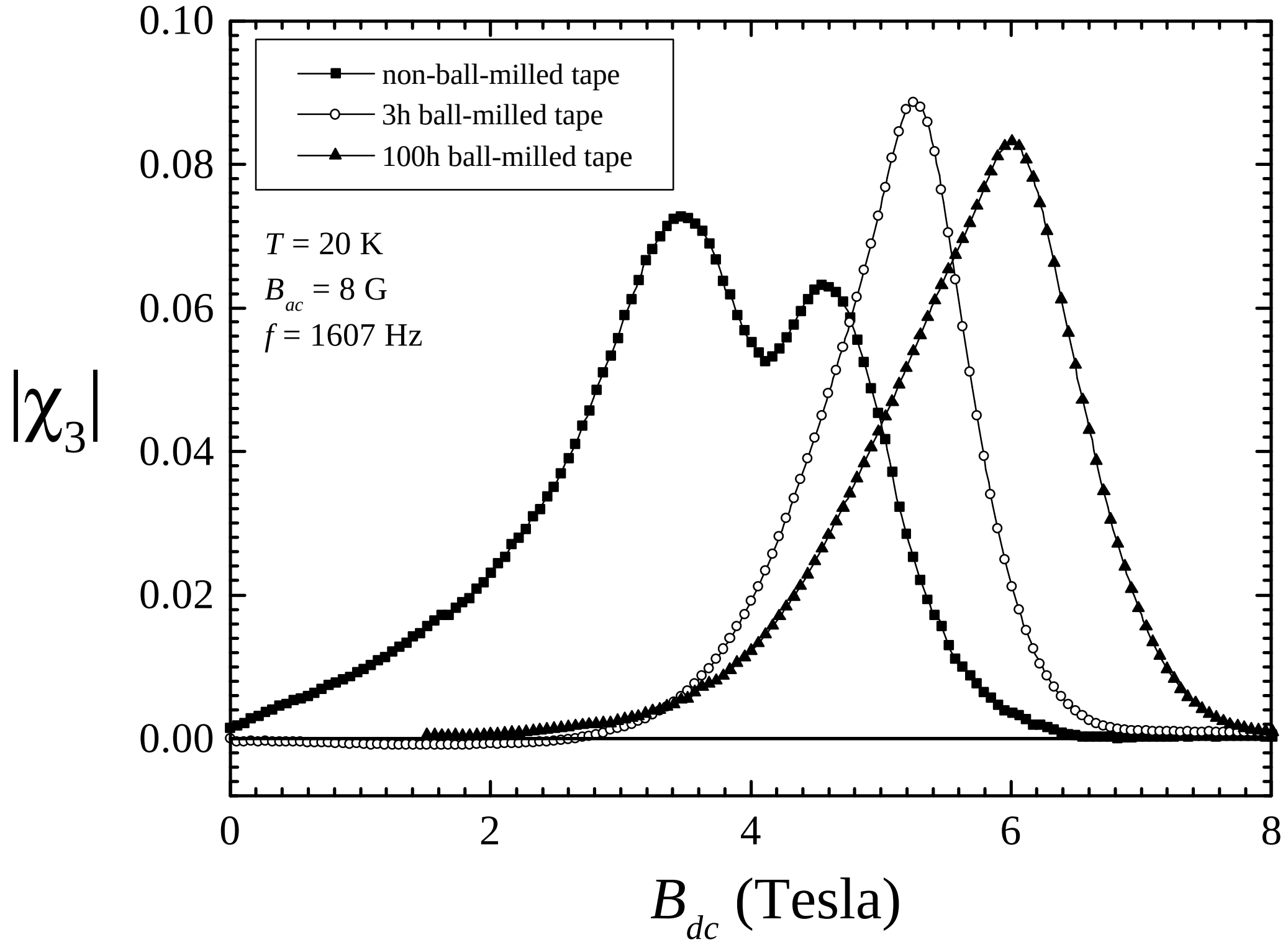

0.10
0.08
0.06
0.04
0.02
0.00
|χ3|
0
2
4
6
8
Bdc (Tesla)
non-ball-milled tape
3h ball-milled tape
100h ball-milled tape
T = 20 K
Bac = 8 G
f = 1607 Hz

Figure 7

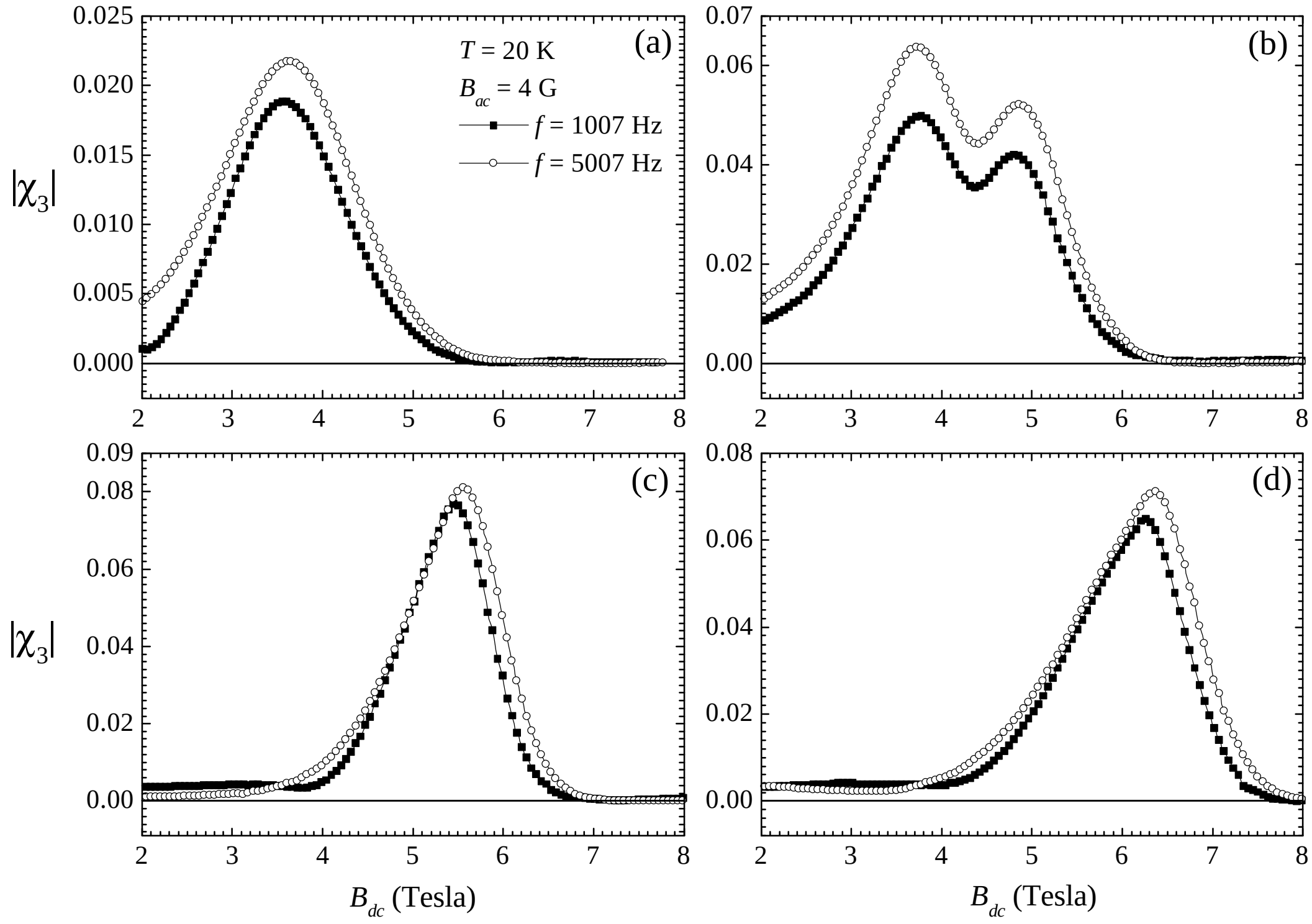

0.025
0.020
0.015
0.010
0.005
0.000
|χ3|
(a)
T = 20 K
Bac = 4 G
f = 1007 Hz
f = 5007 Hz
0.07
0.06
0.04
0.02
0.00
(b)
0.09
0.08
0.06
0.04
0.02
0.00
(c)
0.08
0.06
0.04
0.02
0.00
(d)
2 3 4 5 6 7 8
Bdc (Tesla)

Figure 8

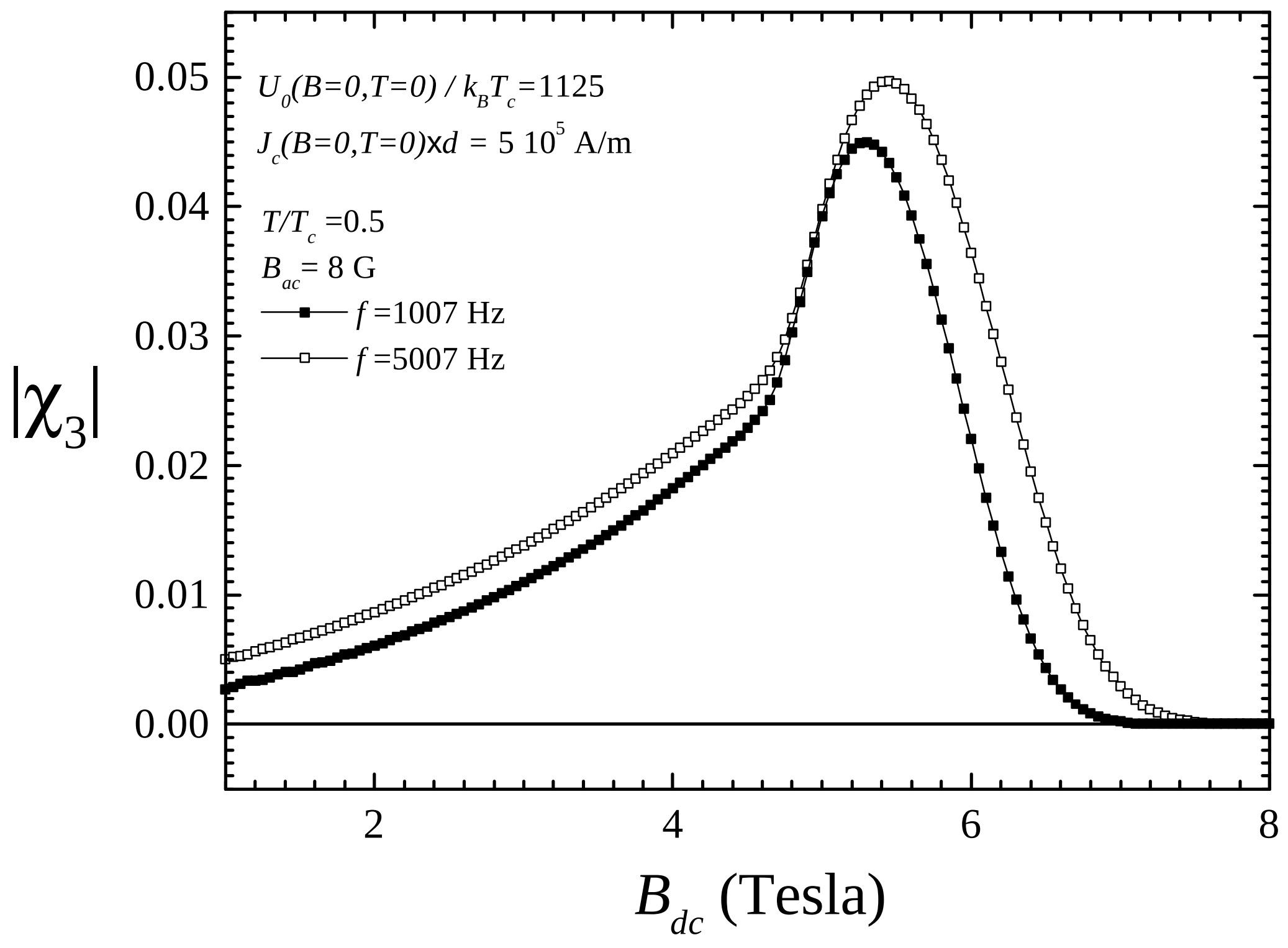

U0(B=0,T=0) / kBTc=1125
Jc(B=0,T=0)xd = 5 10^5 A/m
T/Tc =0.5
Bac= 8 G
f =1007 Hz
f =5007 Hz
|χ3|
0.05
0.04
0.03
0.02
0.01
0.00
2
4
6
8
Bdc (Tesla)

Figure 9